\documentclass[aps,prl,reprint,groupedaddress,showpacs]{revtex4-1} 
 
\usepackage{graphicx}% Include figure files
\usepackage{dcolumn}% Align table columns on decimal point
\usepackage{bm}% bold math
\begin{document} 
 
% Use the \preprint command to place your local institutional report 
% number in the upper righthand corner of the title page in preprint mode. 
% Multiple \preprint commands are allowed. 
% Use the 'preprintnumbers' class option to override journal defaults 
% to display numbers if necessary 
%\preprint{} 
 
%Title of paper 
\title{Microbunching Instability Based Synchrotron Radiation Sources} 
 
% repeat the \author .. \affiliation  etc. as needed 
% \email, \thanks, \homepage, \altaffiliation all apply to the current 
% author. Explanatory text should go in the []'s, actual e-mail 
% address or url should go in the {}'s for \email and \homepage. 
% Please use the appropriate macro foreach each type of information 
 
% \affiliation command applies to all authors since the last 
% \affiliation command. The \affiliation command should follow the 
% other information 
% \affiliation can be followed by \email, \homepage, \thanks as well. 
\author{Atoosa Meseck} 
%\email[]{atoosa.meseck@helmholtz-berlin.de} 
%\homepage[]{Your web page} 
%\thanks{} 
%\altaffiliation{} 
\affiliation{Helmholtz-Zentrum Berlin, Germany} 
 
%Collaboration name if desired (requires use of superscriptaddress 
%option in \documentclass). \noaffiliation is required (may also be 
%used with the \author command). 
%\collaboration can be followed by \email, \homepage, \thanks as well. 
%\collaboration{} 
%\noaffiliation 
 
\date{\today} 
 
\begin{abstract} 
Although microbunching-instability (MBI) based synchrotron-radiation
facilities do not exist yet, the prominent presence of this space-charge
driven instability in the high-brightness accelerator facilities promotes the
ongoing worldwide effort to mastering it. Thereby not only the possible
countermeasures but also t      he applications of MBI are in focus of the
research. So far, the observed microbunching wavelengths are in IR or optical
regime. Shifting the microbunching towards shorter wavelengths seems to
require either very high-bunch peak currents or extremely small
emittances. These extreme characteristics are difficult to generate and hard to
preserve. This letter proposes a novel magnetic device which allows to amplify
the initial shot noise  at wavelengths in EUX and XUV range with moderate peak
current and emittance. It consists of permanent magnets. The specific
combination of focusing channels, dispersive chicanes and single focusing
elements in this device allows for optimal amplification of microbunching by
the longitudinal-space charge-impedance. Extensive simulation studies for
different beam energies show the performance of the proposed device, which fed for example with an energy recovery linac can provide ultra-short coherent-high power-radiation pulses with a high repetition rate.     
% insert abstract here 
\end{abstract} 
 
% insert suggested PACS numbers in braces on next line 
% insert suggested keywords - APS authors don't need to do this 
 
%\maketitle must follow title, authors, abstract, \pacs, and \keywords 
\maketitle 
 
% body of paper here - Use proper section commands 
% References should be done using the \cite, \ref, and \label commands 
\section{} 
 
Similar to the free-electron laser (FEL) instability, the MBI launches out the
initial shot noise in the bunch but in contrast to the FEL instability its
gain, i.e. the amplification of the microbunching due to the MBI, is
not dependent on the radiation field in the undulator. On the one hand, the MBI
can cause a significant degradation of the beam quality and is thus a serious
threat for example to the performance of an FEL facility. On the other hand as the broadband gain of the MBI is not constrained to the resonant wavelength of the undulator it provides a very attractive alternative for the generation of intense broadband radiation. Accordingly, MBI, possible countermeasures and applications are subject of ongoing research worldwide \cite{borland,SaldIIa, heifets, zhiroIIa, zhiroIIb, shaftIa, marcoI, SaldIIb, danielI, gover1, henrik, marcoII, agostino, yurkoschneid, agostinoII, liti}.

In order to build a radiation source based on the space-charge driven MBI
requires besides a driver accelerator providing a high brightness electron beam a
section where the MBI can take place. The beam cross section and
divergence as well as the longitudinal dispersion need to be adjusted and
controlled along this dedicated section to drive the
instability in a controlled manner for a given wavelength range. Throughout
the letter, this section will be referred to as the microbunching section. A
short undulator tuned to the desired wavelength follows the microbunching section,
allowing the microbunched beam to emit high-power radiation. The core piece of
such a facility is the microbunching section.  The requirements and performance of this section determine the requests on the driver accelerator and the length of the undulator. To identify its requirements and to optimize its performance, one has to examine the emergence of the space-charge driven-microbunching and its amplification. 
%\section{Emergence and Amplification of Space Charge Driven Microbunching Instability} 
 
The MBI has its origin in the longitudinal-space charge-forces caused by
density fluctuations in a relativistic high brightness electron beam. The
density fluctuations - i.e. the initial shot noise in the bunch - couple to
the bunch's own impedance leading to a longitudinal electrical field (self
field), which gives rise to a broadband energy modulation within the same
bunch. Dependent on its respective value, the longitudinal dispersion in the
accelerator converts this energy modulation into a density modulation,
amplifying some of the initial fluctuations while suppressing others. Thus,
density modulations within a certain wavelength range are amplified. The
process is called microbunching as the beam becomes microbunched at the
wavelengths for which the amplitude of the fluctuations are enhanced.  
 
Of course the microbunching is accompanied by an increase of the  potential
(space-charge) energy of the bunch which needs to be compensated for by the
available kinetic energy to obey the energy conservation. For a relativistic
bunch passing through an accelerator, the amount of the available kinetic
energy depends on the accelerator element. For example in a field free
section, i.e. a drift, the available kinetic energy is given by the energy spread of the bunch, while in a magnetic chicane the available energy is amplified proportional to the bending angle squared allowing for a much stronger enhancement of the density modulation \cite{liti}.  
The amplified
density modulation in turn leads to a stronger space-charge
impedance, causing more energy modulation and thus amplifying further the
density modulation in the preferred wavelengths. This way, the space-charge impedance within the bunch drives the microbunching instability. 
 
To design and optimize the microbunching section for wavelengths in EUV and XUV range,
one first needs to assess the amount of initial density fluctuations at these
wavelengths. For the sake of readability the wavelengths in EUV and XUV range
will be referred to as EUV-wavelengths throughout this document.
Due to the finite number of electrons in the bunch, there is a significant
amount of noise in transverse and longitudinal electron distribution.
The simplified model of the space-charge force -- which is frequently used to
investigate the microbunching instability in FEL facilities -- focuses only on
the on-axis longitudinal space-charge field utilizing a one-dimensional model
(1D-model)\cite{borland, SaldIIa, heifets, zhiroIIa, zhiroIIb, shaftIa,
  marcoI}. It comprises only the
longitudinal fluctuations, leading to a longitudinal electrical field which
has its maximum value around a modulation wavelength of $\lambda_{1D} = 4\pi
r_b / \gamma_0$, where $r_b$ is the transverse size of the cylindrical bunch, $\gamma_0$ is
the average relativistic factor. For wavelengths shorter than $\lambda_{1D}$
the field decays with the wavelength squared. 

However, transverse density
fluctuations can translate into significant fluctuations in the longitudinal
electrical field with wavelengths much shorter than $\lambda_{1D}$
\cite{marcoI, SaldIIb, danielI, gover1, henrik, marcoII}. Taking the
transverse fluctuations into account, significant density fluctuations for EUV-wavelengths can be expected
\cite{marcoII}. The effect of the transverse fluctuations starts to become
significant for wavelengths with $\lambda_{3D} \leq  r_b / 0.26 \gamma_0 $
whereby the amplitudes of these fluctuations tend to a constant value
independent of the respective wavelength \cite{marcoII}.

Space-charge induced MBI at wavelengths shorter than predicted by 1D-model has
already been observed \cite{zhiroIIb, shaftIa, SaldIIb}
confirming thus the validity of the prediction of 3D-model for the density
fluctuation at short wavelengths. However, for a dedicated generation of
microbunching at a desired EUV-wavelength the existence of corresponding
density fluctuations does not suffice. Additionally, a suitable impedance needs to be
produced and maintained along the respective microbunching section. A suitable impedance is the impedance which
generates the ``right'' amplitude of the energy modulation leading to the 
maximum microbunching at the desired wavelength range in connection with the
longitudinal dispersion, $r_{56}$, of the section. 

In the 3D-model the effects of the transverse RMS beam size, $\sigma_{tra}$, and
divergence, $\sigma_{tra'}$, on the microbunching are taken into
account. Calculating the bunching factor for a given wavenumber, k, observed
at small angles ($\theta_x, \theta_y$) relative to the longitudinal direction,
one obtains for the ensemble average in the 3D-model \cite{danielI}:
 %-------    
\begin{center}
\[
<|b_c(k)|^2> \approx \frac{4}{3N}\left ( \frac{ I_0 r_{56} L_d}{\gamma_0 I_A\sigma_{tra}^2} \right
)^2 \]
\begin{equation}
\times \frac{\exp{\left( -\frac{k^2 r_{56}^2 \sigma_E^2}{2\gamma_0^2} -\frac{k^2 (R_2^2+\theta_y^2 r_{34}^2) \sigma_{tra'}^2}{2} \right)}}{1+\gamma_0^2(R_1^2+\theta_y^2 r_{33}^2)},  
\label{bunching}
\end{equation}
\end{center}
%-------
where $I_0$ is the bunch peak current, $I_A=17\,kA$ is the Alfven current, $L_d$ is
the length of the microbunching section, $\sigma_E$ is the energy spread,
$R_1= r_{51}+\theta_x r_{11}$, $R_2= r_{52}+\theta_x r_{12}$, and $r_{ij}$ are
the elements of the transfer matrix of the microbunching section. The coherent microbunching described in the above formula generates coherent radiation of power:
%-------
\begin{center}
\begin{equation}
P_{coh} \approx \frac{I_0^2}{4\pi\,\sigma_{tra}^2} \frac{K^2 L_u^2}{8 \gamma_0^2} <|b_c(k)|^2> ,
\label{bpower}
\end{equation}
\end{center}
%-------
in a helical undulator of the length $L_u$, where $Z_0$=377\,$\Omega$ is the free
space impedance and K is the undulator parameter.  
%-------  
\begin{figure} 
\includegraphics*[width=75mm]{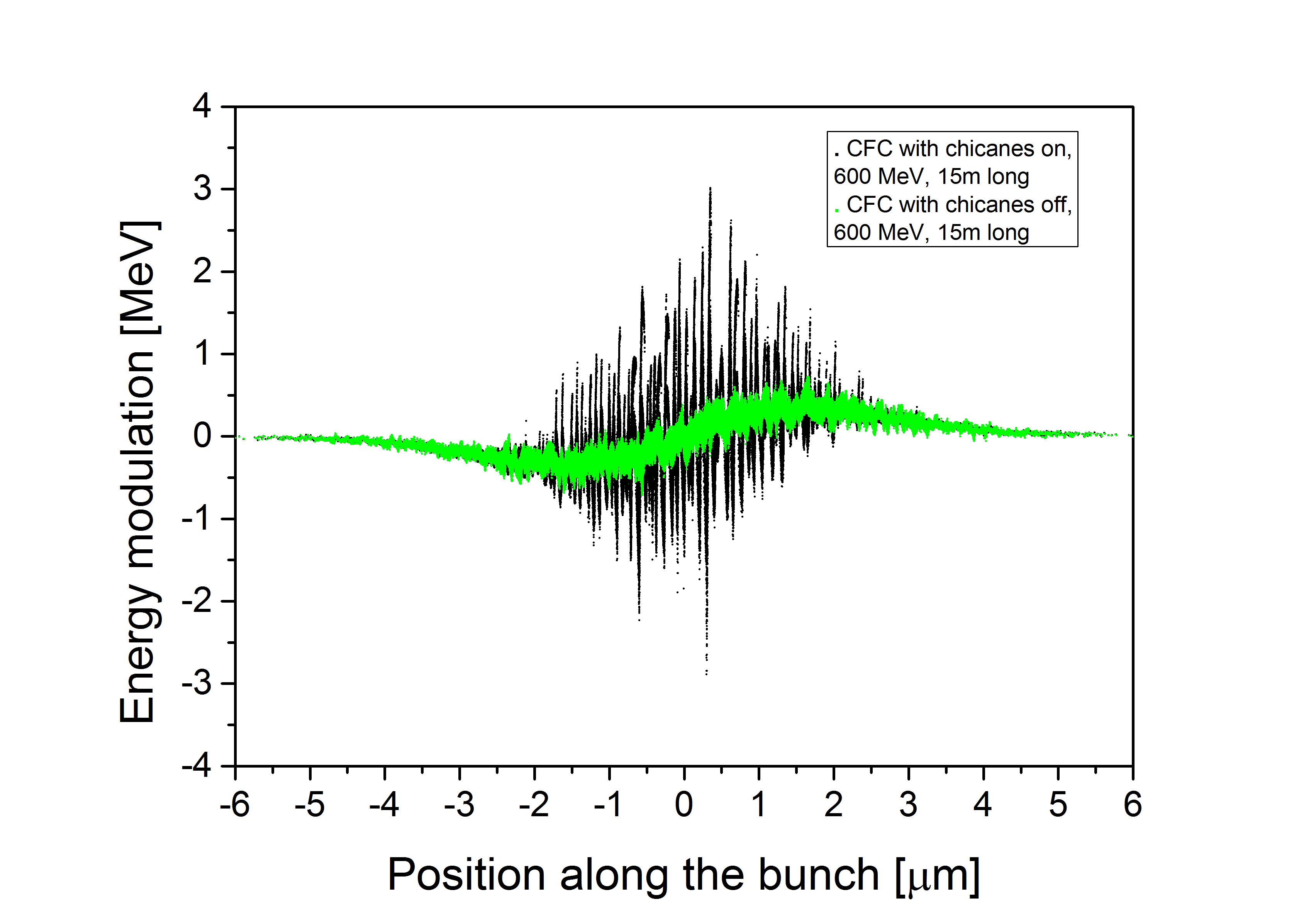} 
\caption{The MBI-caused energy modulations of 600\,MeV bunches in a 15~\,m long section of compact focusing channel (see
Fig.~\ref{fi-CFC-mag-schematic2}) with magnetic chicanes switched on (black
dots) and off (green dots).} 
\label{fi-availkin-case}
\end{figure}
%-------
For the maximum bunching, the exponential function in the Eq.~\ref{bunching}
has to be adjusted to a value of nearly one. Assuming that the magnetic structure
of the microbunching section is such that the two terms
in the argument of the exponential function are independent, suitable values
for $r_{56}$ and $R_2$ can be determined in dependency of the wavelength
range, the energy spread and the transverse beam divergence. 
In principle a FODO lattice with the length L
is an example for such a structure,
where $r_{56}=L/\gamma_0^2$ is independent of $R_2$. 

However, as a significant enhancement of
microbunching requires an enhancement of the available kinetic energy of the
bunch \cite{liti}, 
the microbunching section needs to contain magnetic chicanes, which shovel
effectively a part of the longitudinal momentum of the bunch
into the transverse momenta. Figure~\ref{fi-availkin-case} illustrates the
beneficial impact of magnetic chicanes on the energy modulation. 
Unfortunately magnetic chicanes couple the transverse and longitudinal
motion of the particles and thus the respective elements of the transfer 
matrix, so that in their case $r_{56}$ and $R_2$ are not independent anymore.
%-------
\begin{figure} 
\includegraphics*[width=80mm]{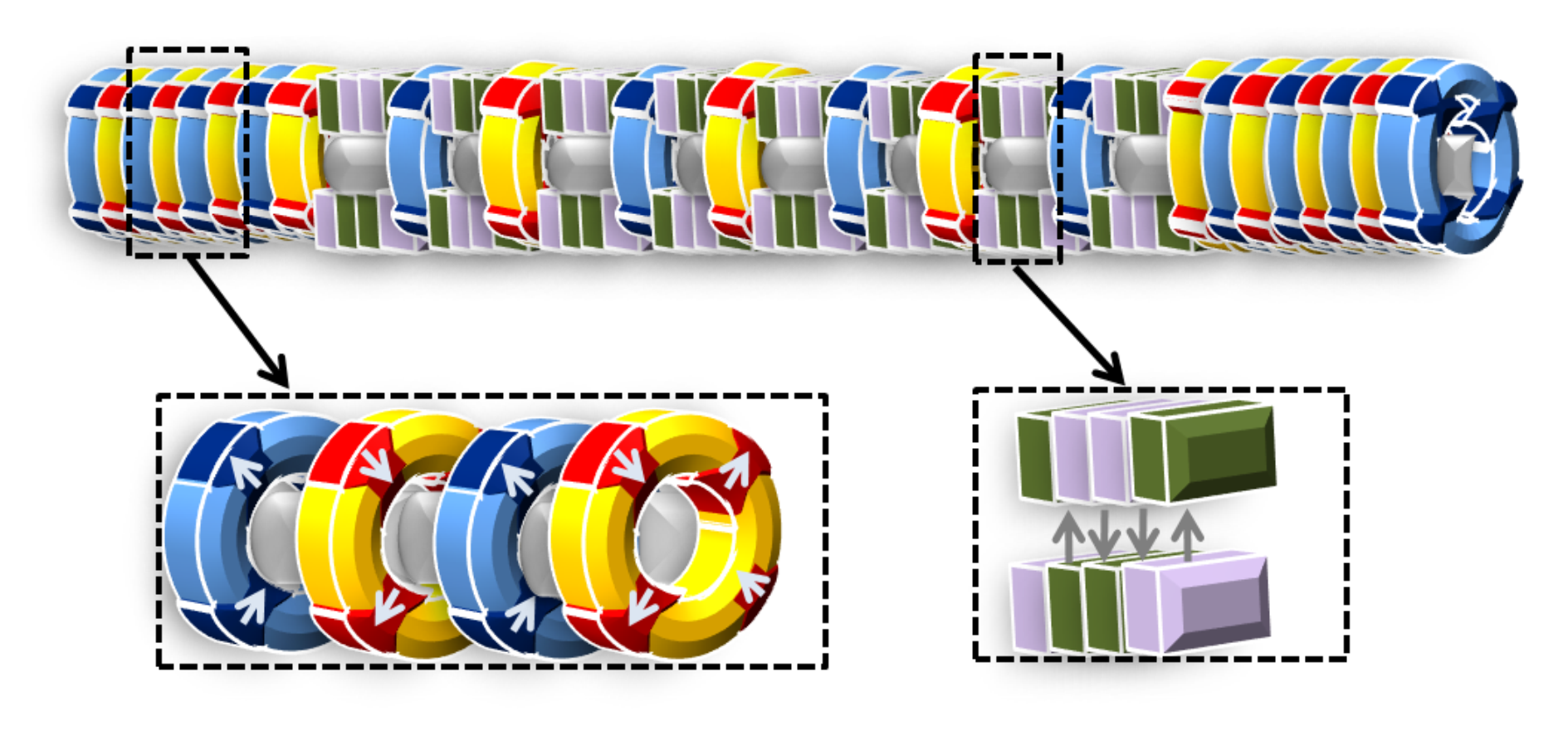} 
\caption{Schematic view of a CFC piece. The device consists of short permanent
  magnet quadrupoles (dark blue-blue and red-yellow rings)and dipoles
  (green-violet blocks), building a combination of pure
  FODO structures and C-chicane intersections.} 
\label{fi-CFC-mag-schematic2}
\end{figure}
%-------
 
Considering the above discussion the value of the exponential
function in the Eq.~\ref{bunching} can at best be optimized by adjusting the
$r_{56}$ using a magnetic chicane and reducing simultaneously the transverse
divergence, $\sigma_{tra'}$, by reducing the focusing strength in the section
which of course leads to an increase of the beam size, $\sigma_{tra},$ for a
given transverse emittances. Once the $r_{56}$ and $\sigma_{tra}$ are fixed,
the values of 
$I_0$, $\gamma_0$ and $L_d$ can be adjusted to maximize the first multiplicand
in the Eq.~\ref{bunching} in order to maximize the bunching factor. 
Generally, the beam energy is not a freely-selectable variable because on one hand the longitudinal-space charge-impedance,
which drives the MBI, decreases with an increasing beam energy, for biding too high
values of $\gamma_0$. On the other hand the beam energy also needs
to satisfy the resonance condition of the undulator for a desired
EUV-wavelength, thus at least for conventional undulators also low values for $\gamma_0$ are unfavorable. 
The length $L_d$ of the microbunching section is also restricted to a few
meters, as $r_{56}=L_d/\gamma_0^2$ of a longer section smears out the energy
modulation before it has a chance to lead to further microbunching
amplification in a chicane. 
Therefore, high-bunch peak currents in combination with very small emittances
and energy spreads, i.e. a very high beam brightness seems to be
required to drive the MBI. In particular for EUV-wavelengths, this
imposes extremely high demands on the accelerator facility. 
%-------
\begin{figure} 
\includegraphics*[width=60mm]{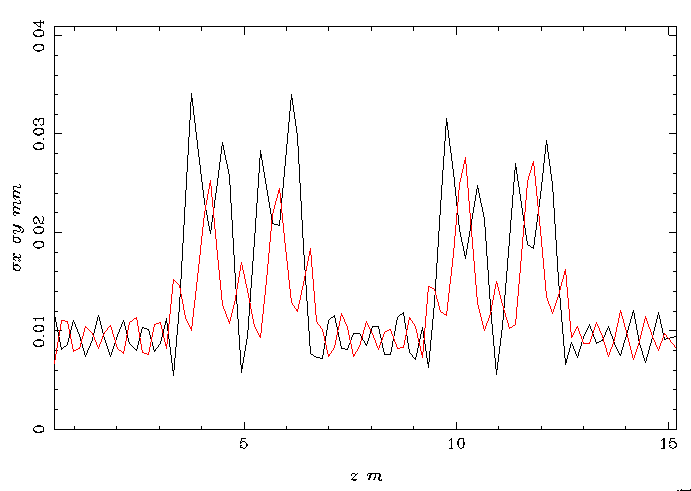} 
\includegraphics*[width=60mm]{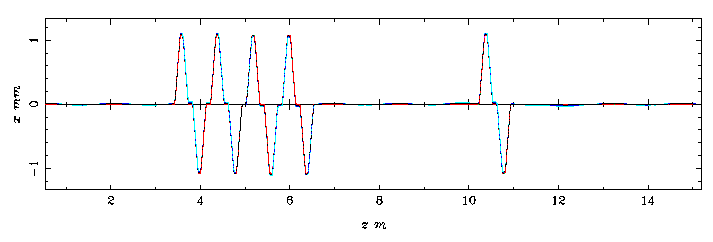} 
\caption{An example for the development of the transverse beam sizes (top) and
  beam trajectory (bottom)  along a 15\,m long section of CFC optimized for
  EUV-wavelengths is shown.} 
\label{fi-tranj}
\end{figure}  
%-------  
  
However, the extreme demands on the accelerator facility can significantly be relaxed
by means of a novel magnetic device, which combines in an very compact manner sections of strong-short quadrupole magnets, arranged in dense FODO-structure, with sections of short dipole magnets, arranged in C-chicane structure. The device, called compact focusing channel (CFC), consists
of permanent-magnet quadrupoles -- for example such as described in
\cite{mpq1, mpq2} with a length of 2\,-5\,cm -- and permanent-magnet
dipoles of a length of 1\,-2\,cm,  providing focusing and
bunching in the same device. An schematic view of the CFC is shown
in Fig.~\ref{fi-CFC-mag-schematic2}. Similar to undulators also
CFCs can be designed for different wavelength ranges, whereby the length of
the FODO-structure and chicane sections, the strength of the quadrupoles and
dipoles, and the number of magnetic chicanes in the CFC are the
properties which have to be designed and optimized. Also the impact
of the coherent synchrotron radiation (CSR) \cite{esaldin} on the microbunched beam
has to be taken into account for an optimal design of the chicane sequence. A
fine tuning of the quadrupole and dipole strengths of an existing CFC is possible for
example by moving the magnets towards the beam axis or further out
\cite{johannes}. 
Figure~\ref{fi-tranj} shows the development of the beam size and beam
trajectory along a 15\,m long section of a CFC optimized for the EUV-wavelengths. 
While the FODO-structure 
provides a small average beam size of about 10\,$\mu$m, a sequence of chicanes provide the
necessary $r_{56}$ of about a few hundred $\mu$m. Thereby, the quadrupoles inside the chicane sequence
ensure that the $\sigma_{tra'}$ is small in this section. 
The CFC has a circular aperture of
8\,mm in this case.  

%---------------
\begin{figure*} 
\includegraphics*[width=130mm]{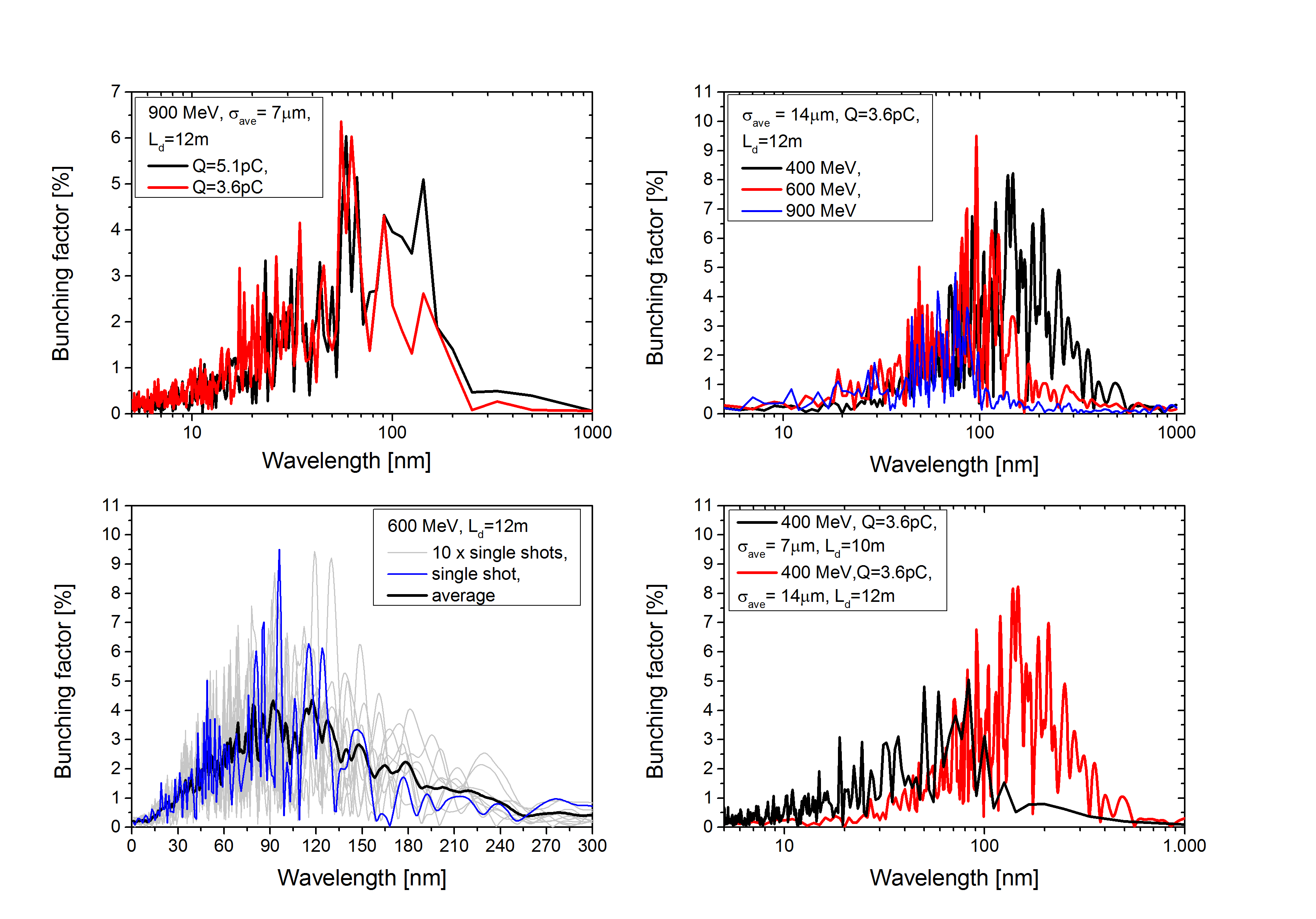} 
\caption{Bunching factor as a function of wavelength is shown for
  different beam properties and CFC settings. The bunching factor for 10
  consecutive shots is shown in the lower-left graph. For these simulations
  the average beam size in the CFC was adjusted to 14\,$\mu$m. A reduction of the
  transverse beam size shifts the bunching towards shorter EUV-wavelengths
  but it also can decreases the maximum value of the bunching as shown in lower-right graph. Also an increase of the bunch charge does not automatically
  increase the bunching factor for short 
wavelengths as shown in upper-left graph for 900\,MeV bunches. The impact of the beam energy of the
bunching factor is depicted in the upper-right graph.}
\label{fi-bunching-all} 
\end{figure*} 
%---------------
To investigate the performance of the CFC for EUV-wavelengths extensive
simulation studies have been carried out utilizing the 3D-space charge-code
ASTRA~\cite{astra}. In the simulations electron bunches with two different
charges of 3.6\,pC and 5.1\,pC, three different beam energies of 400\,MeV, 600\,MeV
and 900\,MeV, two different emittances of 0.1\,mm\,mrad and 0.4\,mm\,mrad
-- leading to different average transverse beam sizes of 14\,$\mu$m and
7\,$\mu$m --, and two 
energy spreads of 10\,keV and 15\,keV (for 900\,MeV cases) with always the same
RMS bunch length of 1.5\,$\mu$m are modeled using always 200000 macroparticle
generated by ASTRA generator. The bunches are 
tracked trough a 15\,m long CFC where the 3D-FFT algorithm of ASTRA with a
dense grid of 1024 longitudinal and 64 transverse grid cells is used.  Generally, the usage of
macroparticles reduces the CPU requirements, however it also leads to an artificially
increased initial shot noise amplitude, so that the CFC length needed for
a desired bunching factor is shorter than in a real facility. This can easily be
amended by lengthening the device suitably. Using formulas presented in
\cite{esaldin}, the increase of the energy spread due to the CSR can be
calculated. Due to the very low bunch charges, the relative energy deviation
generated by CSR in the chicane sections is on the order of $10^{-5}$ and thus
 almost two orders of magnitude smaller than the generated energy modulation
 on the order of  $10^{-3}$ (see Fig.~\ref{fi-availkin-case}).

Figure~\ref{fi-bunching-all} illustrates the main findings of the simulation studies. As expected starting from shot noise, the MBI shows statistical
characteristics clearly visible in the fluctuation of the bunching factor (
Fig.~\ref{fi-bunching-all} lower-left graph). A reduction of the transverse beam
size shifts not only the bunching towards shorter wavelengths but also decreases the value of the exponential function in
Eq.~\ref{bunching} so that in spite of the increase in the first multiplicand
of the equation the bunching factor itself is lower for the smaller beam size as shown for 400\,MeV
bunches in Fig.~\ref{fi-bunching-all} ( lower-right graph). Also an increase of the
bunch charge does not automatically increase the bunching factor for EUV-wavelengths, as it also drives an increase in the longitudinal beam size which
can dominate leading thus to an enhancement of the bunching for longer
wavelengths as shown in Fig.~\ref{fi-bunching-all} (upper-left
graph). Generally, the beam energy and bunch charge are powerful knobs to
adjust the bunching factor for an existing CFC. The impact of the beam energy of the
bunching factor is depicted in Fig.~\ref{fi-bunching-all} (upper-right
graph). An increase of beam energy shifts the bunching towards shorter
wavelength but it also decreases the amplitude of the bunching factor.    
%--------------- 
\begin{figure} 
\includegraphics*[width=80mm]{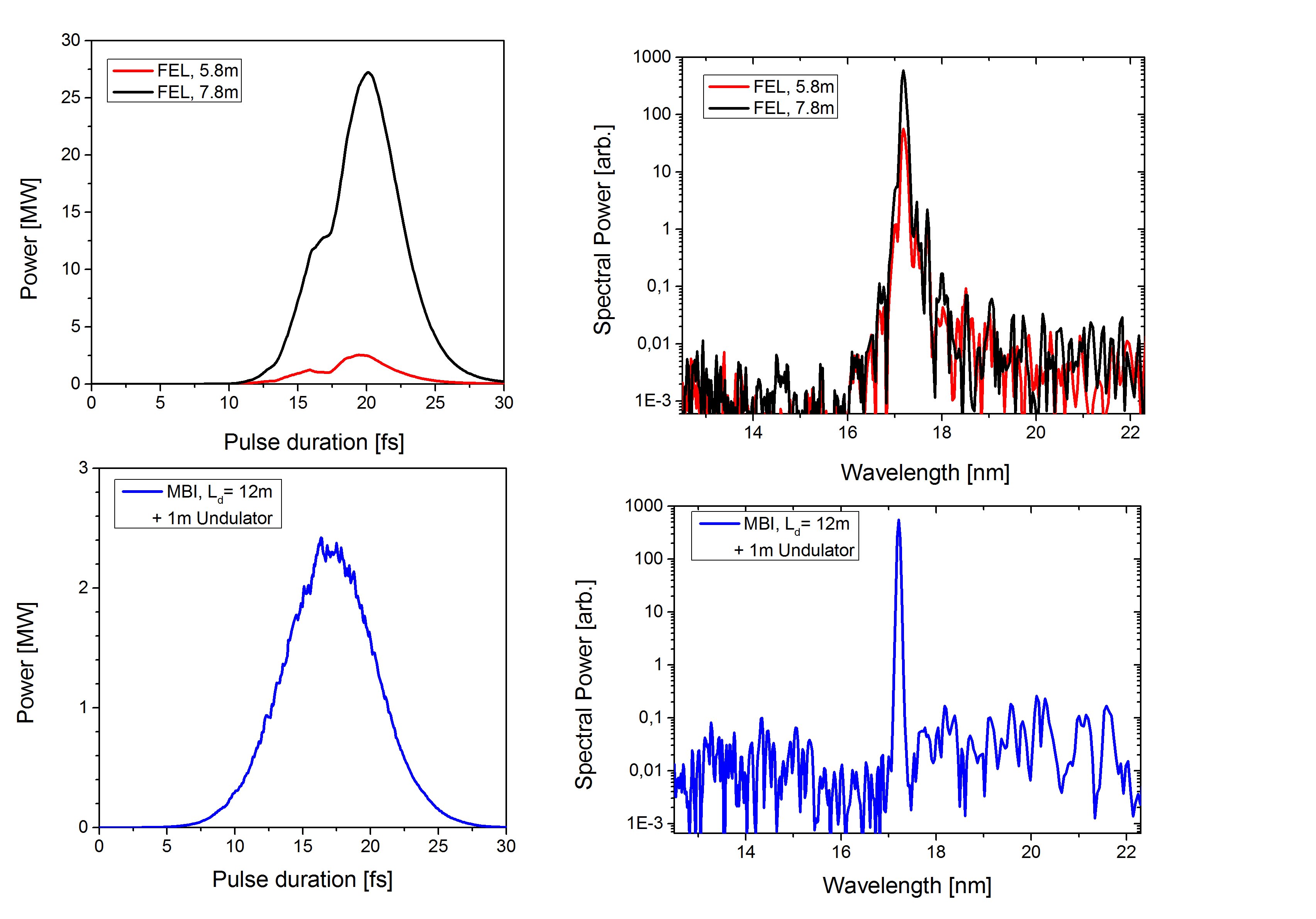} 
\caption{Comparison of simulated radiation power and spectral power profile
  for an FEL and a MBI microbunched undulator output. }
\label{fi-FEL-MBI2} 
\end{figure} 
%---------------

From Eq.~\ref{bpower} one expects high-coherent radiation-power
generated by a microbunched beam in a 1\,m long helical undulator. In order to
verify this expectation and to study the spectral properties of the radiation
generated by a MBI based source, the output of a 900\,MeV ASTRA-run
(7\,$\mu$m-3.6\,PC case in Fig.~\ref{fi-bunching-all} upper-left graph) is used
as an input for coherent radiation simulation, where the 3D-FEL
code GENESIS \cite{gene} is utilized. For comparison also FEL
simulations with the initial, i.e. pre-microbunching, bunch are
performed. Figure~\ref{fi-FEL-MBI2} shows the comparison. 
The MBI-based radiation source has a bandwidth smaller than the FEL source.
After 5.8\,m the FEL generates radiation with the same peak power as the MBI
source in 1\,m \footnote{Of course for a calculation of total length and cost,
  the 12 m CFC needed to produce the bunching has be taken into account.}. After 7.8\,m the FEL reaches the same spectral power. The slight
shoulder in the FEL power profile (Fig.~\ref{fi-FEL-MBI2} (upper-left graph))
is caused by a combination of the slippage and the very short bunch length. It
translates into additional spectral width, limiting the minimum spectral
bandwidth of the FEL for bunch lengths shorter than slippage. The pulse
generated by a MBI based source does not suffer from this effect and thus shows a Fourier transform limited bandwidth. 

CFC devices proposed in this letter allows one to drive the MBI in a desired
wavelength range most efficiently. Thus, they have the potential to be the
core piece of MBI-based synchrotron-radiation sources. Simulation studies
presented prove that utilizing a CFC fed by
a high brightness accelerator and followed by a short undulator, short
(femtosecond long) radiation pulses in EUV range with spectral properties
better than the FEL can be generated. Using CFCs in high repetition rate
accelerators (with GHz rate) allow for the generation of high average power by the subsequent undulator. The combination
of high brightness, high repetition rate and low bunch charge hints in
direction of energy recovery linacs as most the suitable accelerators for
driving CFC based sources. The proposed CFC devices can also amplify a
microbunching imprinted to the bunch by seeding in a prior undulator, thus it can also assist the FEL process. Furthermore, CFCs can be envisaged as microbunching-amplifiers in electron cooling schemes at colliders \cite{daniel3}.    
      
\begin{acknowledgments} 
The author would like to thank A. Gaupp and R. Mitzner for their insightful comments.
% put your acknowledgments here. 
\end{acknowledgments} 

\bibliography{CFC} 

%merlin.mbs apsrev4-1.bst 2010-07-25 4.21a (PWD, AO, DPC) hacked
%Control: key (0)
%Control: author (8) initials jnrlst
%Control: editor formatted (1) identically to author
%Control: production of article title (-1) disabled
%Control: page (0) single
%Control: year (1) truncated
%Control: production of eprint (0) enabled
\providecommand{\noopsort}[1]{}\providecommand{\singleletter}[1]{#1}%
\begin{thebibliography}{24}%
\makeatletter
\providecommand \@ifxundefined [1]{%
 \@ifx{#1\undefined}
}%
\providecommand \@ifnum [1]{%
 \ifnum #1\expandafter \@firstoftwo
 \else \expandafter \@secondoftwo
 \fi
}%
\providecommand \@ifx [1]{%
 \ifx #1\expandafter \@firstoftwo
 \else \expandafter \@secondoftwo
 \fi
}%
\providecommand \natexlab [1]{#1}%
\providecommand \enquote  [1]{``#1''}%
\providecommand \bibnamefont  [1]{#1}%
\providecommand \bibfnamefont [1]{#1}%
\providecommand \citenamefont [1]{#1}%
\providecommand \href@noop [0]{\@secondoftwo}%
\providecommand \href [0]{\begingroup \@sanitize@url \@href}%
\providecommand \@href[1]{\@@startlink{#1}\@@href}%
\providecommand \@@href[1]{\endgroup#1\@@endlink}%
\providecommand \@sanitize@url [0]{\catcode `\\12\catcode `\$12\catcode
  `\&12\catcode `\#12\catcode `\^12\catcode `\_12\catcode `\%12\relax}%
\providecommand \@@startlink[1]{}%
\providecommand \@@endlink[0]{}%
\providecommand \url  [0]{\begingroup\@sanitize@url \@url }%
\providecommand \@url [1]{\endgroup\@href {#1}{\urlprefix }}%
\providecommand \urlprefix  [0]{URL }%
\providecommand \Eprint [0]{\href }%
\providecommand \doibase [0]{http://dx.doi.org/}%
\providecommand \selectlanguage [0]{\@gobble}%
\providecommand \bibinfo  [0]{\@secondoftwo}%
\providecommand \bibfield  [0]{\@secondoftwo}%
\providecommand \translation [1]{[#1]}%
\providecommand \BibitemOpen [0]{}%
\providecommand \bibitemStop [0]{}%
\providecommand \bibitemNoStop [0]{.\EOS\space}%
\providecommand \EOS [0]{\spacefactor3000\relax}%
\providecommand \BibitemShut  [1]{\csname bibitem#1\endcsname}%
\let\auto@bib@innerbib\@empty
%</preamble>
\bibitem [{\citenamefont {Borland}\ \emph {et~al.}(2002)\citenamefont
  {Borland}, \citenamefont {Chae}, \citenamefont {Emma}, \citenamefont
  {Lewellen}, \citenamefont {Bharadwaj}, \citenamefont {Fawley}, \citenamefont
  {Krejcik}, \citenamefont {Limborg}, \citenamefont {Milton}, \citenamefont
  {Nuhn}, \citenamefont {Soliday},\ and\ \citenamefont {Woodley}}]{borland}%
  \BibitemOpen
  \bibfield  {author} {\bibinfo {author} {\bibfnamefont {M.}~\bibnamefont
  {Borland}}, \bibinfo {author} {\bibfnamefont {Y.~C.}\ \bibnamefont {Chae}},
  \bibinfo {author} {\bibfnamefont {P.}~\bibnamefont {Emma}}, \bibinfo {author}
  {\bibfnamefont {J.~W.}\ \bibnamefont {Lewellen}}, \bibinfo {author}
  {\bibfnamefont {V.}~\bibnamefont {Bharadwaj}}, \bibinfo {author}
  {\bibfnamefont {W.~M.}\ \bibnamefont {Fawley}}, \bibinfo {author}
  {\bibfnamefont {P.}~\bibnamefont {Krejcik}}, \bibinfo {author} {\bibfnamefont
  {C.}~\bibnamefont {Limborg}}, \bibinfo {author} {\bibfnamefont {S.~V.}\
  \bibnamefont {Milton}}, \bibinfo {author} {\bibfnamefont {H.-D.}\
  \bibnamefont {Nuhn}}, \bibinfo {author} {\bibfnamefont {R.}~\bibnamefont
  {Soliday}}, \ and\ \bibinfo {author} {\bibfnamefont {M.}~\bibnamefont
  {Woodley}},\ }\href@noop {} {\bibfield  {journal} {\bibinfo  {journal} {Nucl.
  Instrum. Methods Phys. Res. A}\ }\textbf {\bibinfo {volume} {483}},\ \bibinfo
  {pages} {268} (\bibinfo {year} {2002})}\BibitemShut {NoStop}%
\bibitem [{\citenamefont {Saldin}\ \emph {et~al.}(2002)\citenamefont {Saldin},
  \citenamefont {Schneidmiller},\ and\ \citenamefont {Yurkov}}]{SaldIIa}%
  \BibitemOpen
  \bibfield  {author} {\bibinfo {author} {\bibfnamefont {E.~L.}\ \bibnamefont
  {Saldin}}, \bibinfo {author} {\bibfnamefont {E.~A.}\ \bibnamefont
  {Schneidmiller}}, \ and\ \bibinfo {author} {\bibfnamefont {M.~V.}\
  \bibnamefont {Yurkov}},\ }\href@noop {} {\bibfield  {journal} {\bibinfo
  {journal} {Nucl. Instrum. Methods Phys. Res. A}\ }\textbf {\bibinfo {volume}
  {490}},\ \bibinfo {pages} {1} (\bibinfo {year} {2002})}\BibitemShut {NoStop}%
\bibitem [{\citenamefont {Heifets}\ \emph {et~al.}(2002)\citenamefont
  {Heifets}, \citenamefont {Stupakov},\ and\ \citenamefont
  {Krinsky}}]{heifets}%
  \BibitemOpen
  \bibfield  {author} {\bibinfo {author} {\bibfnamefont {S.}~\bibnamefont
  {Heifets}}, \bibinfo {author} {\bibfnamefont {G.}~\bibnamefont {Stupakov}}, \
  and\ \bibinfo {author} {\bibfnamefont {S.}~\bibnamefont {Krinsky}},\ }\href
  {\doibase 10.1103/PhysRevSTAB.5.064401} {\bibfield  {journal} {\bibinfo
  {journal} {Phys. Rev. ST Accel. Beams}\ }\textbf {\bibinfo {volume} {5}},\
  \bibinfo {pages} {064401} (\bibinfo {year} {2002})}\BibitemShut {NoStop}%
\bibitem [{\citenamefont {Huang}\ and\ \citenamefont {Kim}(2002)}]{zhiroIIa}%
  \BibitemOpen
  \bibfield  {author} {\bibinfo {author} {\bibfnamefont {Z.}~\bibnamefont
  {Huang}}\ and\ \bibinfo {author} {\bibfnamefont {K.-J.}\ \bibnamefont
  {Kim}},\ }\href {\doibase 10.1103/PhysRevSTAB.5.074401} {\bibfield  {journal}
  {\bibinfo  {journal} {Phys. Rev. ST Accel. Beams}\ }\textbf {\bibinfo
  {volume} {5}},\ \bibinfo {pages} {074401} (\bibinfo {year}
  {2002})}\BibitemShut {NoStop}%
\bibitem [{\citenamefont {Huang}\ \emph {et~al.}(2004)\citenamefont {Huang},
  \citenamefont {Borland}, \citenamefont {Emma}, \citenamefont {Wu},
  \citenamefont {Limborg}, \citenamefont {Stupakov},\ and\ \citenamefont
  {Welch}}]{zhiroIIb}%
  \BibitemOpen
  \bibfield  {author} {\bibinfo {author} {\bibfnamefont {Z.}~\bibnamefont
  {Huang}}, \bibinfo {author} {\bibfnamefont {M.}~\bibnamefont {Borland}},
  \bibinfo {author} {\bibfnamefont {P.}~\bibnamefont {Emma}}, \bibinfo {author}
  {\bibfnamefont {J.}~\bibnamefont {Wu}}, \bibinfo {author} {\bibfnamefont
  {C.}~\bibnamefont {Limborg}}, \bibinfo {author} {\bibfnamefont
  {G.}~\bibnamefont {Stupakov}}, \ and\ \bibinfo {author} {\bibfnamefont
  {J.}~\bibnamefont {Welch}},\ }\href {\doibase 10.1103/PhysRevSTAB.7.074401}
  {\bibfield  {journal} {\bibinfo  {journal} {Phys. Rev. ST Accel. Beams}\
  }\textbf {\bibinfo {volume} {7}},\ \bibinfo {pages} {074401} (\bibinfo {year}
  {2004})}\BibitemShut {NoStop}%
\bibitem [{\citenamefont {Shaftan}\ and\ \citenamefont
  {Huang}(2004)}]{shaftIa}%
  \BibitemOpen
  \bibfield  {author} {\bibinfo {author} {\bibfnamefont {T.}~\bibnamefont
  {Shaftan}}\ and\ \bibinfo {author} {\bibfnamefont {Z.}~\bibnamefont
  {Huang}},\ }\href {\doibase 10.1103/PhysRevSTAB.7.080702} {\bibfield
  {journal} {\bibinfo  {journal} {Phys. Rev. ST Accel. Beams}\ }\textbf
  {\bibinfo {volume} {7}},\ \bibinfo {pages} {080702} (\bibinfo {year}
  {2004})}\BibitemShut {NoStop}%
\bibitem [{\citenamefont {Venturini}(2007)}]{marcoI}%
  \BibitemOpen
  \bibfield  {author} {\bibinfo {author} {\bibfnamefont {M.}~\bibnamefont
  {Venturini}},\ }\href {\doibase 10.1103/PhysRevSTAB.10.104401} {\bibfield
  {journal} {\bibinfo  {journal} {Phys. Rev. ST Accel. Beams}\ }\textbf
  {\bibinfo {volume} {10}},\ \bibinfo {pages} {104401} (\bibinfo {year}
  {2007})}\BibitemShut {NoStop}%
\bibitem [{\citenamefont {Saldin}\ \emph {et~al.}(2004)\citenamefont {Saldin},
  \citenamefont {Schneidmiller},\ and\ \citenamefont {Yurkov}}]{SaldIIb}%
  \BibitemOpen
  \bibfield  {author} {\bibinfo {author} {\bibfnamefont {E.~L.}\ \bibnamefont
  {Saldin}}, \bibinfo {author} {\bibfnamefont {E.~A.}\ \bibnamefont
  {Schneidmiller}}, \ and\ \bibinfo {author} {\bibfnamefont {M.~V.}\
  \bibnamefont {Yurkov}},\ }\href@noop {} {\bibfield  {journal} {\bibinfo
  {journal} {Nucl. Instrum. Methods Phys. Res. A}\ }\textbf {\bibinfo {volume}
  {528}},\ \bibinfo {pages} {355} (\bibinfo {year} {2004})}\BibitemShut
  {NoStop}%
\bibitem [{\citenamefont {Ratner}\ \emph {et~al.}(2008)\citenamefont {Ratner},
  \citenamefont {Chao},\ and\ \citenamefont {Huang}}]{danielI}%
  \BibitemOpen
  \bibfield  {author} {\bibinfo {author} {\bibfnamefont {D.}~\bibnamefont
  {Ratner}}, \bibinfo {author} {\bibfnamefont {A.}~\bibnamefont {Chao}}, \ and\
  \bibinfo {author} {\bibfnamefont {Z.}~\bibnamefont {Huang}},\ }in\ \href
  {http://accelconf.web.cern.ch/AccelConf/FEL2008/papers/tupph041.pdf} {\emph
  {\bibinfo {booktitle} {FEL Conference}}}\ (\bibinfo {address} {Gyeongju,
  Korea},\ \bibinfo {year} {2008})\ pp.\ \bibinfo {pages}
  {338--341}\BibitemShut {NoStop}%
\bibitem [{\citenamefont {Gover}\ and\ \citenamefont {Dyunin}(2009)}]{gover1}%
  \BibitemOpen
  \bibfield  {author} {\bibinfo {author} {\bibfnamefont {A.}~\bibnamefont
  {Gover}}\ and\ \bibinfo {author} {\bibfnamefont {E.}~\bibnamefont {Dyunin}},\
  }\href {\doibase 10.1103/PhysRevLett.102.154801} {\bibfield  {journal}
  {\bibinfo  {journal} {Phys. Rev. Lett.}\ }\textbf {\bibinfo {volume} {102}},\
  \bibinfo {pages} {154801} (\bibinfo {year} {2009})}\BibitemShut {NoStop}%
\bibitem [{\citenamefont {Loos}\ \emph {et~al.}(2008)\citenamefont {Loos},
  \citenamefont {Akre}, \citenamefont {Brachmann}, \citenamefont {Decker},
  \citenamefont {Ding}, \citenamefont {Dowell}, \citenamefont {Emma},
  \citenamefont {Frisch}, \citenamefont {Gilevich}, \citenamefont {Hays},
  \citenamefont {Hering}, \citenamefont {Huang}, \citenamefont {Iverson},
  \citenamefont {Limborg-Deprey}, \citenamefont {Miahnahri}, \citenamefont
  {Molloy}, \citenamefont {Nuhn}, \citenamefont {Turner}, \citenamefont
  {Welch}, \citenamefont {White}, \citenamefont {Wu},\ and\ \citenamefont
  {Ratner}}]{henrik}%
  \BibitemOpen
  \bibfield  {author} {\bibinfo {author} {\bibfnamefont {H.}~\bibnamefont
  {Loos}}, \bibinfo {author} {\bibfnamefont {R.}~\bibnamefont {Akre}}, \bibinfo
  {author} {\bibfnamefont {A.}~\bibnamefont {Brachmann}}, \bibinfo {author}
  {\bibfnamefont {F.-J.}\ \bibnamefont {Decker}}, \bibinfo {author}
  {\bibfnamefont {Y.~T.}\ \bibnamefont {Ding}}, \bibinfo {author}
  {\bibfnamefont {D.}~\bibnamefont {Dowell}}, \bibinfo {author} {\bibfnamefont
  {P.}~\bibnamefont {Emma}}, \bibinfo {author} {\bibfnamefont {J.~C.}\
  \bibnamefont {Frisch}}, \bibinfo {author} {\bibfnamefont {A.}~\bibnamefont
  {Gilevich}}, \bibinfo {author} {\bibfnamefont {G.~R.}\ \bibnamefont {Hays}},
  \bibinfo {author} {\bibfnamefont {P.}~\bibnamefont {Hering}}, \bibinfo
  {author} {\bibfnamefont {Z.}~\bibnamefont {Huang}}, \bibinfo {author}
  {\bibfnamefont {R.~H.}\ \bibnamefont {Iverson}}, \bibinfo {author}
  {\bibfnamefont {C.}~\bibnamefont {Limborg-Deprey}}, \bibinfo {author}
  {\bibfnamefont {A.}~\bibnamefont {Miahnahri}}, \bibinfo {author}
  {\bibfnamefont {S.}~\bibnamefont {Molloy}}, \bibinfo {author} {\bibfnamefont
  {H.-D.}\ \bibnamefont {Nuhn}}, \bibinfo {author} {\bibfnamefont {J.~L.}\
  \bibnamefont {Turner}}, \bibinfo {author} {\bibfnamefont {J.~J.}\
  \bibnamefont {Welch}}, \bibinfo {author} {\bibfnamefont {W.~E.}\ \bibnamefont
  {White}}, \bibinfo {author} {\bibfnamefont {J.}~\bibnamefont {Wu}}, \ and\
  \bibinfo {author} {\bibfnamefont {D.~F.}\ \bibnamefont {Ratner}},\ }in\ \href
  {http://accelconf.web.cern.ch/AccelConf/FEL2008/papers/thbau01.pdf} {\emph
  {\bibinfo {booktitle} {FEL Conference}}}\ (\bibinfo {address} {Gyeongju,
  Korea},\ \bibinfo {year} {2008})\ pp.\ \bibinfo {pages}
  {485--489}\BibitemShut {NoStop}%
\bibitem [{\citenamefont {Venturini}(2008)}]{marcoII}%
  \BibitemOpen
  \bibfield  {author} {\bibinfo {author} {\bibfnamefont {M.}~\bibnamefont
  {Venturini}},\ }\href {\doibase 10.1103/PhysRevSTAB.11.034401} {\bibfield
  {journal} {\bibinfo  {journal} {Phys. Rev. ST Accel. Beams}\ }\textbf
  {\bibinfo {volume} {11}},\ \bibinfo {pages} {034401} (\bibinfo {year}
  {2008})}\BibitemShut {NoStop}%
\bibitem [{\citenamefont {Marinelli}\ and\ \citenamefont
  {Rosenzweig}(2010)}]{agostino}%
  \BibitemOpen
  \bibfield  {author} {\bibinfo {author} {\bibfnamefont {A.}~\bibnamefont
  {Marinelli}}\ and\ \bibinfo {author} {\bibfnamefont {J.~B.}\ \bibnamefont
  {Rosenzweig}},\ }\href {\doibase 10.1103/PhysRevSTAB.13.110703} {\bibfield
  {journal} {\bibinfo  {journal} {Phys. Rev. ST Accel. Beams}\ }\textbf
  {\bibinfo {volume} {13}},\ \bibinfo {pages} {110703} (\bibinfo {year}
  {2010})}\BibitemShut {NoStop}%
\bibitem [{\citenamefont {Schneidmiller}\ and\ \citenamefont
  {Yurkov}(2010)}]{yurkoschneid}%
  \BibitemOpen
  \bibfield  {author} {\bibinfo {author} {\bibfnamefont {E.~A.}\ \bibnamefont
  {Schneidmiller}}\ and\ \bibinfo {author} {\bibfnamefont {M.~V.}\ \bibnamefont
  {Yurkov}},\ }\href {\doibase 10.1103/PhysRevSTAB.13.110701} {\bibfield
  {journal} {\bibinfo  {journal} {Phys. Rev. ST Accel. Beams}\ }\textbf
  {\bibinfo {volume} {13}},\ \bibinfo {pages} {110701} (\bibinfo {year}
  {2010})}\BibitemShut {NoStop}%
\bibitem [{\citenamefont {Marinelli}\ \emph {et~al.}(2013)\citenamefont
  {Marinelli}, \citenamefont {Hemsing}, \citenamefont {Dunning}, \citenamefont
  {Xiang}, \citenamefont {Weathersby}, \citenamefont {O'Shea}, \citenamefont
  {Gadjev}, \citenamefont {Hast},\ and\ \citenamefont
  {Rosenzweig}}]{agostinoII}%
  \BibitemOpen
  \bibfield  {author} {\bibinfo {author} {\bibfnamefont {A.}~\bibnamefont
  {Marinelli}}, \bibinfo {author} {\bibfnamefont {E.}~\bibnamefont {Hemsing}},
  \bibinfo {author} {\bibfnamefont {M.}~\bibnamefont {Dunning}}, \bibinfo
  {author} {\bibfnamefont {D.}~\bibnamefont {Xiang}}, \bibinfo {author}
  {\bibfnamefont {S.}~\bibnamefont {Weathersby}}, \bibinfo {author}
  {\bibfnamefont {F.}~\bibnamefont {O'Shea}}, \bibinfo {author} {\bibfnamefont
  {I.}~\bibnamefont {Gadjev}}, \bibinfo {author} {\bibfnamefont
  {C.}~\bibnamefont {Hast}}, \ and\ \bibinfo {author} {\bibfnamefont {J.~B.}\
  \bibnamefont {Rosenzweig}},\ }\href {\doibase 10.1103/PhysRevLett.110.264802}
  {\bibfield  {journal} {\bibinfo  {journal} {Phys. Rev. Lett.}\ }\textbf
  {\bibinfo {volume} {110}},\ \bibinfo {pages} {264802} (\bibinfo {year}
  {2013})}\BibitemShut {NoStop}%
\bibitem [{\citenamefont {Litvinenko}\ and\ \citenamefont {Wang}(2014)}]{liti}%
  \BibitemOpen
  \bibfield  {author} {\bibinfo {author} {\bibfnamefont {V.}~\bibnamefont
  {Litvinenko}}\ and\ \bibinfo {author} {\bibfnamefont {G.}~\bibnamefont
  {Wang}},\ }in\ \href@noop {} {\emph {\bibinfo {booktitle} {FEL Conference}}}\
  (\bibinfo {address} {BASEL, Swiss},\ \bibinfo {year} {2014})\BibitemShut
  {NoStop}%
\bibitem [{\citenamefont {Eichner}\ \emph {et~al.}(2007)\citenamefont
  {Eichner}, \citenamefont {Gr\"uner}, \citenamefont {Becker}, \citenamefont
  {Fuchs}, \citenamefont {Habs}, \citenamefont {Weingartner}, \citenamefont
  {Schramm}, \citenamefont {Backe}, \citenamefont {Kunz},\ and\ \citenamefont
  {Lauth}}]{mpq1}%
  \BibitemOpen
  \bibfield  {author} {\bibinfo {author} {\bibfnamefont {T.}~\bibnamefont
  {Eichner}}, \bibinfo {author} {\bibfnamefont {F.}~\bibnamefont {Gr\"uner}},
  \bibinfo {author} {\bibfnamefont {S.}~\bibnamefont {Becker}}, \bibinfo
  {author} {\bibfnamefont {M.}~\bibnamefont {Fuchs}}, \bibinfo {author}
  {\bibfnamefont {D.}~\bibnamefont {Habs}}, \bibinfo {author} {\bibfnamefont
  {R.}~\bibnamefont {Weingartner}}, \bibinfo {author} {\bibfnamefont
  {U.}~\bibnamefont {Schramm}}, \bibinfo {author} {\bibfnamefont
  {H.}~\bibnamefont {Backe}}, \bibinfo {author} {\bibfnamefont
  {P.}~\bibnamefont {Kunz}}, \ and\ \bibinfo {author} {\bibfnamefont
  {W.}~\bibnamefont {Lauth}},\ }\href {\doibase 10.1103/PhysRevSTAB.10.082401}
  {\bibfield  {journal} {\bibinfo  {journal} {Phys. Rev. ST Accel. Beams}\
  }\textbf {\bibinfo {volume} {10}},\ \bibinfo {pages} {082401} (\bibinfo
  {year} {2007})}\BibitemShut {NoStop}%
\bibitem [{\citenamefont {Becker}\ \emph {et~al.}(2009)\citenamefont {Becker},
  \citenamefont {Bussmann}, \citenamefont {Raith}, \citenamefont {Fuchs},
  \citenamefont {Weingartner}, \citenamefont {Kunz}, \citenamefont {Lauth},
  \citenamefont {Schramm}, \citenamefont {El~Ghazaly}, \citenamefont
  {Gr\"uner}, \citenamefont {Backe},\ and\ \citenamefont {Habs}}]{mpq2}%
  \BibitemOpen
  \bibfield  {author} {\bibinfo {author} {\bibfnamefont {S.}~\bibnamefont
  {Becker}}, \bibinfo {author} {\bibfnamefont {M.}~\bibnamefont {Bussmann}},
  \bibinfo {author} {\bibfnamefont {S.}~\bibnamefont {Raith}}, \bibinfo
  {author} {\bibfnamefont {M.}~\bibnamefont {Fuchs}}, \bibinfo {author}
  {\bibfnamefont {R.}~\bibnamefont {Weingartner}}, \bibinfo {author}
  {\bibfnamefont {P.}~\bibnamefont {Kunz}}, \bibinfo {author} {\bibfnamefont
  {W.}~\bibnamefont {Lauth}}, \bibinfo {author} {\bibfnamefont
  {U.}~\bibnamefont {Schramm}}, \bibinfo {author} {\bibfnamefont
  {M.}~\bibnamefont {El~Ghazaly}}, \bibinfo {author} {\bibfnamefont
  {F.}~\bibnamefont {Gr\"uner}}, \bibinfo {author} {\bibfnamefont
  {H.}~\bibnamefont {Backe}}, \ and\ \bibinfo {author} {\bibfnamefont
  {D.}~\bibnamefont {Habs}},\ }\href {\doibase 10.1103/PhysRevSTAB.12.102801}
  {\bibfield  {journal} {\bibinfo  {journal} {Phys. Rev. ST Accel. Beams}\
  }\textbf {\bibinfo {volume} {12}},\ \bibinfo {pages} {102801} (\bibinfo
  {year} {2009})}\BibitemShut {NoStop}%
\bibitem [{\citenamefont {Saldin}\ \emph {et~al.}(1997)\citenamefont {Saldin},
  \citenamefont {Schneidmiller},\ and\ \citenamefont {Yurkov}}]{esaldin}%
  \BibitemOpen
  \bibfield  {author} {\bibinfo {author} {\bibfnamefont {E.~L.}\ \bibnamefont
  {Saldin}}, \bibinfo {author} {\bibfnamefont {E.~A.}\ \bibnamefont
  {Schneidmiller}}, \ and\ \bibinfo {author} {\bibfnamefont {M.~V.}\
  \bibnamefont {Yurkov}},\ }\href {\doibase 10.1016/S0168-9002(97)00822-X}
  {\bibfield  {journal} {\bibinfo  {journal} {Nucl. Instrum. Methods Phys. Res.
  A}\ }\textbf {\bibinfo {volume} {398}},\ \bibinfo {pages} {373} (\bibinfo
  {year} {1997})}\BibitemShut {NoStop}%
\bibitem [{\citenamefont {Bahrdt}(2014)}]{johannes}%
  \BibitemOpen
  \bibfield  {author} {\bibinfo {author} {\bibfnamefont {J.}~\bibnamefont
  {Bahrdt}},\ }\href@noop {} {\enquote {\bibinfo {title} {Private
  communication},}\ }\bibinfo {howpublished} {HZB, Berlin} (\bibinfo {year}
  {2014})\BibitemShut {NoStop}%
\bibitem [{\citenamefont {Fl{\"o}ttmann}(2000)}]{astra}%
  \BibitemOpen
  \bibfield  {author} {\bibinfo {author} {\bibfnamefont {K.}~\bibnamefont
  {Fl{\"o}ttmann}},\ }\href {http://www.desy.de/~mpyflo} {\emph {\bibinfo
  {title} {ASTRA: A Space Charge Tracking Algorithm}}},\ \bibinfo
  {organization} {DESY},\ \bibinfo {address} {Hamburg,
  \path|www.desy.de/~mpyflo|} (\bibinfo {year} {2000})\BibitemShut {NoStop}%
\bibitem [{\citenamefont {Reiche}(1999)}]{gene}%
  \BibitemOpen
  \bibfield  {author} {\bibinfo {author} {\bibfnamefont {S.}~\bibnamefont
  {Reiche}},\ }\href@noop {} {\bibfield  {journal} {\bibinfo  {journal} {Nucl.
  Instrum. Methods Phys. Res. A}\ }\textbf {\bibinfo {volume} {429}},\ \bibinfo
  {pages} {243} (\bibinfo {year} {1999})}\BibitemShut {NoStop}%
\bibitem [{Note1()}]{Note1}%
  \BibitemOpen
  \bibinfo {note} {Of course for a calculation of total length and cost, the 12
  m CFC needed to produce the bunching has be taken into account.}\BibitemShut
  {Stop}%
\bibitem [{\citenamefont {Ratner}(2013)}]{daniel3}%
  \BibitemOpen
  \bibfield  {author} {\bibinfo {author} {\bibfnamefont {D.}~\bibnamefont
  {Ratner}},\ }\href {\doibase 10.1103/PhysRevLett.111.084802} {\bibfield
  {journal} {\bibinfo  {journal} {Phys. Rev. Lett.}\ }\textbf {\bibinfo
  {volume} {111}},\ \bibinfo {pages} {084802} (\bibinfo {year}
  {2013})}\BibitemShut {NoStop}%
\end{thebibliography}%
 
\end{document}